\documentstyle[12pt]{article}
 
  {\bigskip\noindent{\bf Acknowledgements}\bigskip}%
  {\bigskip}
  {\bigskip\noindent{\bf Note added}\bigskip}%
  {\bigskip}
\newcommand{\beqn}{\begin{equation}}
\newcommand{\eeqn}{\end{equation}}
\newcommand{\beqnarray}{\begin{eqnarray}}
\newcommand{\eeqnarray}{\end{eqnarray}}
\newcommand{\rd}{\partial}
\newcommand{\dfrac}[2]{ \frac{\displaystyle #1}{\displaystyle #2} }

\newcommand{\diag}{\mathop{\mbox{\rm diag}}}

\newcommand{\Abar}{{\bar{A}}}

\newcommand{\bC}{{\bf C}}

\newcommand{\cF}{{\cal F}}

\newcommand{\rbar}{{\bar r}}

\newcommand{\tbar}{{\bar{t}}}
\newcommand{\Tbar}{{\bar{T}}}

\begin{document}


\begin{flushright}
hep-th/9603069\\
KUCP-0092\\
March 1996
\end{flushright}
\bigskip

\begin{center}
\begin{Large}
\bf
Isomonodromic Deformations and \\[6pt]
Supersymmetric Gauge Theories
\end{Large}
\bigskip
\bigskip

Kanehisa TAKASAKI \raisebox{2mm}{{\footnotesize 1}{$\dagger$}}
and
Toshio NAKATSU \raisebox{2mm}{{\footnotesize 2}}
\bigskip

\begin{small}
$~^{1}${\it Department of Fundamental Sciences,
            Faculty of Integrated Human Studies, Kyoto University }\\
       {\it Yoshida-Nihonmatsu-cho, Sakyo-ku, Kyoto 606,Japan}\\
$~^{2}${\it Department of Mathematics and Physics, Ritsumeikan University }\\
       {\it Kusatsu, Shiga 525-77, Japan}\\
\end{small}
\end{center}
\bigskip
\bigskip
\bigskip

\begin{center}
\bf ABSTRACT
\end{center}
\bigskip

\begin{small}
\noindent 
Seiberg-Witten solutions of four-dimensional supersymmetric gauge 
theories possess rich but involved integrable structures. The goal 
of this paper is to show that an isomonodromy problem provides a 
unified framework for understanding those various features of 
integrability.  The Seiberg-Witten solution itself can be 
interpreted as a WKB limit of this isomonodromy problem. The 
origin of underlying Whitham dynamics (adiabatic deformation of 
an isospectral problem), too, can be similarly explained by a 
more refined asymptotic method (multiscale analysis).  The case 
of $N=2$ SU($s$) supersymmetric Yang-Mills theory without matter 
is considered in detail for illustration. The isomonodromy problem 
in this case is closely related to the third Painlev\'e equation 
and its multicomponent analogues.  An implicit relation to $t\tbar$ 
fusion of topological sigma models is thereby expected. 
\end{small}

\vfill
\hrule
\vskip 3mm
\begin{small}
\noindent{$\dagger$}
E-mail: takasaki@yukawa.kyoto-u.ac.jp, nakatsu@tkyvax.phys.s.u-tokyo.ac.jp
\end{small}

\newpage


\section{Introduction}

The so called ``Seiberg-Witten solutions'' of 
four-dimensional supersymmetric gauge theories have common 
building blocks --- a family of complex algebraic curves 
(Riemann surfaces) , a meromorphic differential $dS$ 
on these curves, period integrals of $dS$, and a 
prepotential $\cF$ in the sense of special geometry.  
The original Seiberg-Witten theory \cite{bib:Seiberg-Witten} 
deals with SU(2) theories and uses elliptic curves of 
a special form. 

Recent studies on Seiberg-Witten solutions are deeply 
connected with classical integrable systems. Gorsky et al. 
\cite{bib:Gorsky-etal} first discovered this fact in 
the case of $N=2$ SU(2) supersymmetric Yang-Mills theory 
without matter.  According to their interpretation,  
the elliptic curves are the ``spectral curves'' of 
elliptic solutions to the KdV equation, and $dS$ is a 
differential that emerges in the analysis of 
``adiabatic deformations'' (or ``modulation'') of 
those elliptic solutions by the so called ``Whitham 
averaging method'' \cite{bib:Whitham-averaging}. 

The observation of Gorsky et al. was soon extended 
\cite{bib:Martinec-Warner1,bib:Nakatsu-Takasaki} 
to the solutions for other classical gauge groups 
\cite{bib:Argyres-Faraggi,bib:Klemm-etal,%
bib:Danielsson-Sundborg,bib:Brandhuber-Landsteiner}. 
In these cases, hyperelliptic curves of rather special 
types are used in place of the Seiberg-Witten elliptic 
curves. Martinec and Warner \cite{bib:Martinec-Warner1} 
noticed that these hyperelliptic curves are nothing but 
spectral curves of affine (periodic) Toda chain systems. 
The present authors \cite{bib:Nakatsu-Takasaki}, meanwhile, 
argued that affine Toda field equation lies in the heart 
of the generalized Seiberg-Witten solutions and plays the 
role of the KdV equation in the work of Gorsky et al. 
These two observations are not contradictory, but rather 
complementary.  The fact is that the affine Toda chain system 
corresponds to ``finite-band solutions'' of the affine 
Toda field equation, and that adiabatic deformations of 
these finite band solutions give rise to a system of 
Whitham-type. The prepotential $\cF$, in this respect, 
is just a ``quasi-classical $\tau$ function''
\cite{bib:Nakatsu-Takasaki}.  This characterization of 
the prepotential yielded an interesting application 
\cite{bib:Eguchi-Yang} to the problem of broken scale 
invariance of $N=2$ supersymmetric gauge theories with 
matters \cite{bib:Hanany-Oz,bib:Argyres-Shapere-etal}. 

Recently, $N=4$ supersymmetric Yang-Mils theories, too, 
have been treated in the framework of classical integrable 
systems (typically, the elliptic Calogero-Moser systems) 
\cite{bib:Donagi-Witten,bib:Gorsky-Marshakov,%
bib:Martinec-Warner2,bib:Itoyama-Morozov}

These results on integrable structures of supersymmetric 
gauge theories show that each gauge theory is accompanied 
with three different types of classical integrable systems:
\begin{enumerate}
\item A finite dimensional integrable dynamical system 
(affine Toda chain, elliptic Calogero-Moser system, etc.) 
\item A soliton equation (KdV equation, affine Toda fields, etc.) 
\item A Whitham system (of KdV equation, affine Toda fields, etc.)
\end{enumerate}
These integrable systems are linked with each other as follows:
The first integrable system is solvable by algebro-geometric 
methods, and each solution is associated with an algebraic 
curve (spectral curve).  The moduli space of those spectral 
curves is identified with the quantum moduli space of a 
supersymmetric gauge theory.  To be able to undergo modulation, 
these solutions of the first integrable system are embedded 
into the second integrable system as finite-band solutions.  
The slow dynamics on the moduli space can be separated by 
averaging over the fast (quasi-periodic) dynamics of the 
finite-band solutions. The Whitham system is the equations 
of motion of this slow dynamics.

The present understanding of integrable structures in    
Seiberg-Witten solutions is thus considerably messy. In 
particular, the derivation of the Whitham dynamics looks 
very artificial.  The least persuasive point is that 
it cannot explain the origin of adiabatic deformations; 
it simply {\em assumes} that deformation takes place.  
The origin of the Whitham dynamics is discussed by several 
authors in the literature  
\cite{bib:Gorsky-etal,bib:Martinec-Warner2,bib:Itoyama-Morozov}, 
who mostly seek the origin in (semi-classical) quantization 
of the classical integrable systems arising here. We are not 
satisfied with those explanations. The goal of this paper is 
to present a different approach based on the concept of 
isomonodromic deformations. This approach might be eventually 
absorbed into the idea of renormalization groups, but for 
the moment, this seems to be the most clear and tractable 
approach.

Our approach is inspired by a idea of Flaschka and Newell 
\cite{bib:Flaschka-Newell}. They noted, with a fairly general 
(though speculative) reasoning, that isomonodromic deformations 
in ``WKB approximation'' look like modulation of isospectral 
deformations.  As they added, similar observations were already 
made by Boutroux \cite{bib:Boutroux} and Garnier 
\cite{bib:Garnier} early in this century. According to Boutroux, 
for instance, solutions of the first (PI) and second (PII) 
Painlev\'e equations 
\begin{eqnarray}
    (\mbox{PI})  && \dfrac{d^2u}{dx^2} = u^2 - x, 
                                                            \\
    (\mbox{PII}) && \dfrac{d^2u}{dx^2} = u^2 - xu - \nu,  
\end{eqnarray}
behave like a (modulated) elliptic function as $x \to \infty$. 
The same (but more refined) idea is also applied to the string 
equations of two-dimensional quantum gravity 
\cite{bib:Moore1,bib:Novikov,bib:Fucito-etal}. 
We now attempt to apply this idea to Seiberg-Witten solutions.

\section{WKB analysis of isomonodromy problem}

Let us recall a universal formulation of Seiberg-Witten solutions 
due to Itoyama and Morozov \cite{bib:Itoyama-Morozov}. They write 
the algebraic curves $C$ over the quantum moduli spaces as:
\begin{equation}
    \det \Bigl( w - L(z) \Bigr) = 0, 
\end{equation}
where $z$ is a parameter (``spectral parameter'') on another 
algebraic curve $C_0$ (of genus 0 or 1), and $L(z)$ is a finite 
dimensional matrix depending on $z$ that gives the Lax operator 
of an associated integrable dynamical system.  This ``eigenvalue 
equation'' gives a finite covering $C \to C_0$ (``spectral 
covering''). The meromorphic differential $dS$, too, has the 
simple expression 
\begin{equation}
    dS = w dz. 
\end{equation}
In the case of the $N = 2$ SU($s$) Yang-Mills theory without 
matter, for instance, $C_0$ is a Riemann sphere $\bC P^1$, and the 
spectral parameter $z$ is given by the logarithm of the coordinate 
$h$ on $\bC P^1 \setminus \{0,\infty\}$: 
\begin{equation}
    z = \log h. 
\end{equation}
The matrix $L(z)$ is the Lax operator of the affine SU($s$) Toda
chain: 
\begin{eqnarray}
 && L(z) = \left( \begin{array}{lllll}
            b_1     & 1      &        &        & c_s h^{-1} \\
            c_1     & b_2    & \ddots &        &            \\
                    & \ddots & \ddots & \ddots &            \\
                    &        & \ddots & \ddots & 1          \\
            h       &        &        & c_{s-1}& b_s
           \end{array} \right), 
                                                   \nonumber \\
 && b_j = \dfrac{dq_j}{dt}, \quad  c_j = \exp( q_j - q_{j+1}).  
\end{eqnarray}
The eigenvalue equation of $L(z)$ boils down to the simple 
equation  
\begin{equation}
    h - P(w) + \Lambda^{2s} h^{-1} = 0, 
\end{equation}
where $P(w)$ is a polynomial in $w$ of the form 
$w^s + \sum_{j=2}^s u_j w^{s-j}$. This equation can be solved 
for $h$ as
\begin{equation}
    h = \frac{P(w) + y}{2}, \quad 
    y^2 = P(w) - \Lambda^{2s}, 
\end{equation}
thus yielding the familiar hyperelliptic spectral curve of a finite 
periodic Toda chain. 
    This kind of integrable systems generally have a commuting set 
of isospectral flows with a Lax representation of the form 
\begin{equation}
    \dfrac{\rd L(z)}{\rd t_n} = [ P_n(z), L(z) ], \quad 
    \Biggl[ \dfrac{\rd}{\rd t_n} - P_n(z), 
            \dfrac{\rd}{\rd t_m} - P_m(z) \Biggr] = 0, 
\end{equation}
where $P_n(z)$, like $L(z)$ itself, are suitably selected 
matrix-valued meromorphic functions on $C_0$. The associated 
linear problem 
\begin{equation}
    w \psi(z) = L(z) \psi(z), \quad 
    \dfrac{\rd \psi(z)}{\rd t_n} = P_n(z) \psi(z)
\end{equation}
is integrable in the sense of Frobenius, and the vector- 
(or matrix-)valued solution $\psi(z)$ can be characterized 
as a ``Baker-Akhiezer function'' on $C$ \cite{bib:alg-geom}.  
This, conversely, gives an algebro-geometric method for solving 
(and constructing) this kind of integrable dynamical systems 
\cite{bib:alg-geom} as well as more general ``Hitchin systems'' 
\cite{bib:Hitchin}.

Our idea is to replace the above isospectral problem by the 
following isomonodromy problem: 
\begin{equation}
    \epsilon \dfrac{\rd \Psi(z)}{\rd z} = Q(z) \Psi(z), \quad 
    \dfrac{\rd \Psi(z)}{\rd t_n} = P_n(z) \Psi(z). 
\end{equation}
Here $\epsilon$ is a small parameter that plays the role of 
the Planck constant in the subsequent (formal) ``WKB analysis''.  
$\Psi(z)$ is not the same as $\psi(z)$, though connected with 
$\psi(z)$ by a simple relation as we shall show below.  
$Q(z)$ is conceptually the same as $L(z)$, and reduces to 
$L(z)$ in a limit.  To give a nontrivial isomonodromy problem, 
however, $Q(z)$ has to be a more general matrix (see subsequent 
sections). According to the general theory of isomonodromic 
deformations \cite{bib:Flaschka-Newell,bib:Jimbo-Miwa-Ueno}, 
the $t$-flows leave the monodromy data of the first ODE in $z$ 
if and only if the above linear system is integrable in the 
sense of Frobenius, i.e, the zero-curvature equations
\begin{equation}
    \Biggl[ \dfrac{\rd}{\rd t_n} - P_n(z), 
            \epsilon \dfrac{\rd}{\rd z} - Q(z)  \Biggr] = 0, 
    \quad 
    \Biggl[ \dfrac{\rd}{\rd t_n} - P_n(z), 
            \dfrac{\rd}{\rd t_m} - P_m(z) \biggr] = 0
\end{equation}
are satisfied. These zero-curvature equations, thus, gives 
a Lax representation of an isospectral problem.

The idea of Flaschka and Newell \cite{bib:Flaschka-Newell} 
(and of Novikov \cite{bib:Novikov}) is to put $\Psi(z)$ 
in a WKB form (as $\epsilon \to 0$): 
\begin{equation}
    \Psi(z) = 
    \Bigl( \phi(z) +  \sum_{n=1}^\infty \epsilon^n \phi_n(z) \Bigr) 
    \Bigl( \rd^2 S(z)/\rd z^2 \Bigr)^{-1/2} 
    \exp\Bigl( \epsilon^{-1} S(z) \Bigr). 
\end{equation}
In the leading order of this $\epsilon$-expansion, the ODE in $z$ 
of the above isomonodromy problem gives the algebraic equation 
\begin{equation}
    \frac{\rd S(z)}{\rd z} \phi(z) = Q(z) \phi(z) 
\end{equation}
for the ``amplitude'' $\phi(z)$. If we identify 
\begin{equation}
    w = \frac{\rd S(z)}{\rd z}, 
\end{equation}
this algebraic equation gives essentially the same eigenvalue 
problem as in the formulation of Itoyama and Morozov. (The relation 
between $\phi(z)$ and $\psi(z)$ will be clarified in the next section.) 
The last relation can be rewritten 
\begin{equation}
    dS(z) = w dz. 
\end{equation}
Thus the Seiberg-Witten differential $dS$ can be reproduced from 
the above isomonodromy problem.  

Note that $z$ and $w$ now play the role of a coordinate and 
its conjugate momentum. The passage from the isospectral problem 
to the isomonodromy problem is achieved by the substitution 
\begin{equation}
    w \to \epsilon \frac{\rd}{\rd z}. 
\end{equation}
This is a kind of quantization!  Here $\epsilon$ corresponds to 
the Planck constant, and $\Psi(z)$ to a quantum mechanical wave 
function.  Thus, we are now considering a quantum mechanical 
problem in the (one-dimensional) space of spectral parameter. 
This is in contrast with other proposals on the origin of 
Whitham dynamics 
\cite{bib:Gorsky-etal,bib:Martinec-Warner2,bib:Itoyama-Morozov}, 
which consider quantization in an ordinary coordinate space of 
interacting particle systems.

\section{Multiscale analysis of isomonodromy problem}

The next task is to separate the above isomonodromy problem 
into a combination of fast (isospectral) and slow (Whitham) 
dynamics. Conceptually, this is similar to the Born-Oppenheimer 
approximation in quantum mechanics. ``Multiscale analysis'' is 
a powerful perturbative method for studying this kind of problems, 
and particularly popular among researchers of nonlinear waves 
and pattern formations \cite{bib:multiscale}. Let us apply this 
method to our isomonodromy problem.

Following the idea of multiscale analysis, we introduce two 
sets of variables $t$ and $T$ connected by the relation 
\begin{equation}
    \epsilon t_n = T_n, 
\end{equation}
and assume that that all fields $\{ u_\alpha \}$ in the system 
are functions of variables $t$ and $T$ as $u_\alpha = u_\alpha(t,T)$.  
The two sets of time variables represent fast and slow time scales.  
Derivatives of the fields can be written as a sum of contributions 
from these two scales: 
\begin{equation}
    \dfrac{\rd}{\rd t_n} u_\alpha(t,\epsilon t) 
    = \left. \dfrac{\rd u_\alpha(t,T)}{\rd t_n} 
             + \epsilon \dfrac{\rd u_\alpha(t,T)}{\rd T_n} 
      \right|_{T=\epsilon t}.  
\end{equation}
Thus the coefficient matrices $P_n(z)$ and $Q(z)$ are assumed 
to be functions of $(t,T)$, 
\begin{equation}
    P_n(z) = P_n(t,T,z), \  Q(z) = Q(t,T,z). 
\end{equation}
(To emphasize the roles of $t$ and $T$, we write all independent 
variables explicitly.) We now look for a ``wave function'' of the form
\begin{equation}
    \Psi(z) = 
    \Bigl( \phi(t,T,z) + \sum_{n=1}^\infty \epsilon^n \phi_n(t,T,z) \Bigr) 
    \Bigl( \rd^2 S(T,z) / \rd z^2 \Bigr)^{-1/2}
    \exp\Bigl( \epsilon^{-1} S(T,z) \Bigr). 
\end{equation}
Note that $S(T,z)$ is assumed to be $t$-independent. This is 
an essential ansatz in the following calculations.  

Now, from the leading order of $\epsilon$-expansion, $\phi(t,T,z)$ 
and $S(t,T,z)$ turn out to satisfy the equations 
\begin{eqnarray}
    \dfrac{\rd S(T,z)}{\rd z} \phi(t,T,z) &=& Q(t,T,z) \phi(t,T,z), 
                                                                 \\
    \dfrac{\rd \phi(t,T,z)}{\rd t_n} 
    + \dfrac{\rd S(T,z)}{\rd T_n} \phi(t,T,z) 
    &=& P_n(t,T,z) \phi(t,T,z). 
\end{eqnarray}
These equations can be further converted into a more familiar form
\begin{equation}
    w \psi(t,T,z) = Q(t,T,z) \psi(t,T,z), \quad 
    \dfrac{\rd \psi(t,T,z)}{\rd t_n} = P_n(t,T,z) \psi(t,T,z). 
\end{equation}
where we have defined 
\begin{equation}
    w := \frac{\rd S(T,z)}{\rd z}, \quad 
    \psi(z) :=  
    \phi(z) \exp\Biggl( \sum t_n \frac{\rd S(z)}{\rd T_n} \Biggr). 
\end{equation}
This is exactly an isospectral problem!  The associated spectral 
curve $C$ is defined by the eigenvalue equation
\begin{equation}
    \det\Bigl( w - Q(t,T,z)\Bigr) = 0, 
\end{equation} 
which is $t$-independent, but now depends on $T$. Thus $T$ 
enters into the isospectral problem as ``adiabatic parameters''. 
$w$ is now a meromorphic function on $C$, and the phase function 
$S(T,z)$ is to be reproduced as
\begin{equation}
    S(T,z) = \int^z w dz \ (+\ \mbox{function of $T$ only}). 
\end{equation}

One can now use the algebro-geometric method \cite{bib:alg-geom} 
to construct $\psi(z)$ as the (vector-valued) Baker-Akhiezer 
function of a finite-band solution.  In general, such a 
Baker-Akhiezer function takes the form 
\begin{equation}
    \psi(z) = \phi(z) \exp \Bigl( \sum t_n \Omega_n(z) \Bigr), 
\end{equation}
where $\Omega_n$ are primitive functions of Abelian differentials 
$d\Omega_n$ on $C$, 
\begin{equation}
    \Omega_n(z) = \int^z d\Omega_n. 
\end{equation}
The amplitude part $\phi(z)$ are made from several theta functions 
of the form $\theta(\sum t_n \sigma_n + \ldots)$, where $\sigma_n$ 
is a vector $\sigma_n = {}^t(\sigma_{n,1},\ldots,\sigma_{n,g})$ 
of period integrals 
\begin{equation}
    \sigma_{n,j} = \frac{1}{2\pi i} \oint_{\beta_j} d\Omega_n 
\end{equation} 
of $d\Omega_n$.  These theta functions are responsible for the 
quasi-periodic (fast) dynamics of the isospectral problem, which 
is eventually ``averaged out'' and does not contribute to the 
Whitham (slow) dynamics. The main contribution to the Whitham 
dynamics stems from the exponential part. By matching the above 
algebro-geometric expression of $\psi(z)$ with the previous 
multiscale expression, we find that $S(z)$ and $\Omega_n(z)$ 
have to be connected by the equations 
\begin{equation}
    \dfrac{\rd S(z)}{\rd T_n} = \Omega_n(z). 
\end{equation}
In fact, these are the most fundamental equations (or, rather, 
the very definition) of a Whitham system. 

One may also interpret this construction of an asymptotic 
solution of the isomonodromy problem in the language of 
averaging methods.  The original Whitham averaging method 
\cite{bib:Whitham-averaging} will not be very convenient 
for this purpose; Krichever's averaging method 
\cite{bib:Krichever-averaging} is more suited because 
it is formulated in terms of Baker-Akhiezer functions 
(A similar analysis of the string equations of two-dimensional 
quantum gravity is done by Fucito et al. \cite{bib:Fucito-etal}). 

This is, actually, not the end of the story. The above multiscale 
analysis is more or less formal, and have to be justified on a 
rigorous mathematical foundation. Furthermore, this kind of 
analysis of isomonodromy problems is faced with a new difficulty 
not met in isospectral problems --- Stokes phenomena.  Stokes 
phenomena take place in WKB approximation in such a way that 
the approximation is not uniformly valid over the whole 
$z$ plane (or the Riemann surface $C_0$). The $z$ plane has 
to be divided into subdomains in which the WKB approximation 
is uniformly valid; such domains are determined by the 
configuration of ``(anti)Stokes curves''. Because of this new 
situation, as emphasized by Moore in the case of two-dimensional 
quantum gravity \cite{bib:Moore2}, the above analysis has to 
be done more carefully.  Rigorous mathematical treatment of 
all these issues is left for future researches.

\section{Example: N = 2 SU(s) supersymmetric Yang-Mills theory}

We now illustrate the isomonodromy problem in more detail in 
the case of $N=2$ SU($s$) supersymmetric Yang-Mills theory 
without matter. In particular, the relation between $L(z)$ 
and $Q(z)$, which is left unclear in the previous sections, 
will be clarified.  

The underlying system of soliton equation is the affine 
SU($s$) Toda field hierarchy. This is a periodic reduction 
of the full Toda hierarchy \cite{bib:Ueno-Takasaki}, and 
can also be treated in the language of infinite matrices or 
difference operators, but we here use a more standard 
$s \times s$ matrix formulation. Just like the full Toda 
hierarchy, the SU($s$) version has two infinite series of 
time variables $t = (t_1,t_2,\ldots)$ and 
$\tbar = (\tbar_1,\tbar_2,\ldots)$; a difference is that 
the flows of $t_s,t_{2s},\ldots$ and $\tbar_s,\tbar_{2s},\ldots$ 
are trivial.  All the flows are generated by $s \times s$ 
matrices $A_n(h)$ and $\Abar_n(h)$ and obey the zero-curvature 
equations
\begin{eqnarray}
    \Biggl[ \dfrac{\rd}{\rd t_n} - A_n(h), 
            \dfrac{\rd}{\rd t_m} - A_m(h) \Biggr] &=& 0,  
                                                      \nonumber \\
    \Biggl[ \dfrac{\rd}{\rd \tbar_n} - \Abar_n(h), 
            \dfrac{\rd}{\rd \tbar_m} - \Abar_m(h) \Biggr] &=& 0,  
                                                      \nonumber \\
    \Biggl[ \dfrac{\rd}{\rd t_n} - A_n(h), 
            \dfrac{\rd}{\rd \tbar_m} - \Abar_m(h) \Biggr] &=& 0. 
\end{eqnarray}
The generators $A_1(h)$ and $\Abar_1(h)$ are directly related with 
the field variables $u_j$ of the SU($s$) Toda field equations
\begin{equation}
    \dfrac{\rd^2 u_j}{\rd t_1 \rd \tbar_1} 
    + \exp( u_{j+1} - u_j ) - \exp( u_j - u_{j-1} ) = 0 \quad 
    (\sum_{j=1}^s  u_j = 0, \ u_{j+s} = u_j)
\end{equation}
as: 
\begin{eqnarray}
    A_1 = 
          \left( \begin{array}{llll}
            b_1  &  1      &        & 0     \\
                 &  \ddots & \ddots &       \\
                 &         & \ddots & 1     \\
            h    &         &        & b_s
          \end{array} \right), 
    &&
    \Abar_1 = 
          \left( \begin{array}{llll}
            0    &         &        & c_s h^{-1} \\
            c_1  &  \ddots &        &       \\
                 &  \ddots & \ddots &       \\
            0    &         & c_{s-1}& 0
          \end{array} \right), 
                                                    \nonumber \\
    b_j = \dfrac{\rd u_j}{\rd t_1}, 
    &&
    c_j = \exp( u_{j+1} - u_j ). 
\end{eqnarray}
The generators $A_n(h)$ and $\Abar_n(h)$ of higher flows can be 
written 
\begin{eqnarray}
    A_n(h) &=& 
    \Lambda(h)^n + [\diag]\Lambda(h)^{s-1} + \cdots + [\diag], 
                                                    \nonumber \\
    \Abar(h) 
    &=& [\diag]\Lambda(h)^{-s} + \ldots + [\diag]\Lambda(h)^{-1}, 
\end{eqnarray}
where ``$[\diag]$'' stands for a diagonal matrix independent of 
$h$, and $\Lambda(h)^{\pm}$ are $s \times s$ matrices of the form 
\begin{equation}
    \Lambda(h) = 
          \left( \begin{array}{llll}
            0    &  1      &        & 0     \\
                 &  \ddots & \ddots &       \\
                 &         & \ddots & 1     \\
            h    &         &        & 0
          \end{array} \right), 
    \quad
    \Lambda(h)^{-1} = 
          \left( \begin{array}{llll}
            0    &         &        & h^{-1} \\
            1    &  \ddots &        &       \\
                 &  \ddots & \ddots &       \\
            0    &         & 1      & 0
          \end{array} \right). 
\end{equation}
In particular, the generators of the trivial flows are given by 
powers of $\Lambda(h)$: 
\begin{eqnarray}
   && A_s(h) = \Lambda(h), A_{2s}(h) = \Lambda(h)^2, \ldots, 
                                                     \nonumber \\
   && \Abar_s(h) = \Lambda(h)^{-1}, \Abar_{s}(h) = \Lambda(h)^{-2}, 
    \ldots. 
\end{eqnarray}
As in the full Toda hierarchy \cite{bib:Ueno-Takasaki}, 
$A_n(h)$ and $\Abar_n(h)$ can be expressed more explicitly 
in terms of two extra Lax operators. (Details are omitted here, 
because we do not need them in the following.) The associated 
linear problem is given by 
\begin{equation}
    \dfrac{\rd \Psi}{\rd t_n} = A_n(h) \Psi, \quad 
    \dfrac{\rd \Psi}{\rd \tbar_n} = \Abar_n(h) \Psi, \quad 
    \Psi = {}^t (\psi_1,\ldots,\psi_s). 
\end{equation}
The vector elements  $\psi_1,\ldots,\psi_s$  are, in fact, 
part of the Baker-Akhiezer functions $\psi_j(t,\tbar,\lambda)$ 
($\lambda = h^{1/s}$) of the full Toda hierarchy. In the 
$s$-periodic reduction, they are related as 
\begin{equation}
    \psi_{j+s} = \lambda^s \psi_j = h \psi_j. 
\end{equation}
(In this sense, $h$ is nothing but the Floquet multiplier in 
the spectral analysis of periodic Toda chains.)

We now impose the following homogeneity condition to $\psi_j$'s: 
\begin{equation}
    \lambda \dfrac{\rd \psi_j}{\rd \lambda} 
    = \Biggl( \sum_{n=1}^\infty n t_n \dfrac{\rd}{\rd t_n} 
        + j - \sum_{n=1}^\infty n \tbar_n \dfrac{\rd}{\rd \tbar_n}
      \Biggr) \psi_j.
\end{equation}
This amounts to assigning the following dimensions (or ``weights'') 
to the time and spectral variables, and requiring $\psi_j$ to be a 
homogeneous function of degree $-j$:
\begin{equation}
    t_n \to n, \ \tbar_n \to -n, \ \lambda \to -1. 
\end{equation}
This forces the matrix elements of $A_n(h)$ and $\Abar_n(h)$ 
to be also homogeneous. In particular, $b_j$, $c_j$ and $u_j$ 
become homogeneous functions of degree $-1$, $0$ and $0$.

It is well known that this kind of homogeneity constraints convert 
an isospectral problem (soliton equation) to an isomonodromy 
problem \cite{bib:Flaschka-Newell,bib:Jimbo-Miwa-Ueno}. 
To see how this mechanism works in the present case, let us write 
the homogeneity conditions in the following matrix form:
\begin{equation}
    \lambda \frac{\rd \Psi}{\rd \lambda} 
    = \Biggl( \sum_{n=1}^\infty n t_n \dfrac{\rd}{\rd t_n} 
        + \Delta - \sum_{n=1}^\infty n \tbar_n \dfrac{\rd}{\rd \tbar_n}
      \Biggr) \Psi, 
    \quad 
    \Delta := ( i \delta_{ij} ). 
\end{equation}
One can use the aforementioned linear differential equations to 
rewrite the $t$- and $\tbar$-derivatives on the right hand side. 
This eventually leads to a linear ODE of the form 
\begin{equation}
    \epsilon h \frac{\rd \Psi}{\rd h} = 
    \epsilon \frac{\rd \Psi}{\rd z} = Q(z) \Psi, 
\end{equation}
where 
\begin{equation}
    Q(z) 
    := \frac{\epsilon}{s} \Biggl( \sum_{n=1}^\infty n t_n A_n(h)
        + \Delta - \sum_{n=1}^\infty n \tbar_n \Abar_n(h) \Biggr). 
\end{equation}
This is exactly the monodromy problem that we have sought for!

More precisely, this ``constrained'' affine Toda field hierarchy 
is not yet an isomonodromy problem. A true isomonodromy problem 
arises when only a finite number of time variables are left nonzero. 
For instance, given two positive integers $r$ and $\rbar$ , the 
constrained hierarchy restricted to the finite dimensional submanifold
\begin{eqnarray}
    t_r = \frac{s}{\epsilon r}, && 
    t_{r+1} = t_{r+2} = \ldots = 0, 
                                                         \nonumber \\
    \tbar_\rbar = - \frac{s}{\epsilon \rbar}, &&
    \tbar_{\rbar+1} = \tbar_{\rbar+2} = \ldots = 0, 
\end{eqnarray}
gives an isomonodromy problem. The discrete parameters $(r,\rbar)$ 
resemble the index of a ``critical point'' or a ``universality class'' 
in two-dimensional quantum gravity. Hopefully, our isomonodromy 
problems, too, might have some ``stringy'' interpretation.

The case of $(r,\rbar) = (1,1)$ is special, because no free time 
variables are left here. It is, however, exactly in this case that 
the Lax operator $L(z)$ of the Seiberg-Witten solution is reproduced:  
\begin{equation}
    \lim_{\epsilon \to 0} \left. Q(z) 
      \right|_{t_1=-\tbar_1=s/\epsilon,t_2=\tbar_2=\ldots=0} 
    = A_1(h) + \Abar_1(h) 
    = L(z). 
\end{equation}
(Note that the term proportional to $\Delta$ simply disappears 
in the limit as $\epsilon$.) Thus, in a strict sense, the 
Seiberg-Witten solution itself corresponds to no isospectral 
problem.  The remaining flows in the Seiberg-Witten solution 
are flows in the Seiberg-Witten periods
\begin{equation}
     a_j = \oint_{\alpha_j} dS
\end{equation}
or in their dual periods $a_{Dj}$; they are however not 
isomonodromic. The fact is that a tower of isomonodromic 
families with different indices $(r,\rbar)$  are ``hidden'' 
behind the Seiberg-Witten solution, and they become visible 
only when embedded into the affine Toda hierarchy.

An interesting situation takes place if $t_1$ and $\tbar_1$ 
are free variables while $t_2 = t_3 = \ldots = 0$ and 
$\tbar_2 = \tbar_3 = \ldots = 0$. In this setting, the 
homogeneity condition implies that the Toda field variables 
$u_j$ are functions of the radial coordinate
\begin{equation}
    x := 4 \sqrt{t_1 \tbar_1}
\end{equation}
only. (The numerical factor ``$4$'' is inserted to make the 
final answer simpler.) The Toda field equations now becomes 
a system of ODE's:
\begin{equation}
    4 \dfrac{d^2 u_j}{dx^2} + 4 \dfrac{1}{x}\dfrac{du_j}{dx}
    + \exp\Bigl( u_{j+1} - u_j \Bigr) 
    - \exp\Bigl( u_j - u_{j-1} \Bigr) = 0. 
\end{equation}
In particular, in the case of SU(2), just a single independent 
field $u$ is left  ($u_1 = -u_2 = u/2$) and obeys an ODE of 
the form
\begin{equation}
    \frac{d^2 u}{dx^2} + \frac{1}{x} \frac{du}{dx} 
    + \sinh u = 0.
\end{equation}
This is an expression of the third Painleve equation (PIII). 
The system in the SU($s$) case may be interpreted as a 
multicomponent analogue of PIII.  Remarkably, these equations 
of PIII type also emerge in $t\tbar$ fusion of topological 
sigma models \cite{bib:Cecotti-Vafa}.  (This kind of 
isomonodromy problems in $t\tbar$ fusion models are considered 
in a more general context by Dubrovin \cite{bib:Dubrovin-ttbar}.)

The affine Toda field hierarchy, thus, plays the same role as 
the KdV and modified KdV hierarchies do in two-dimensional 
quantum gravity. These hierarchies of soliton equations provides 
a universal framework in which to consider different models 
(isomonodromy problems) on an equal footing. The associated 
Whitham hierarchies, too, has a universal structure, as 
demonstrated in our previous paper \cite{bib:Nakatsu-Takasaki}. 
This hierarchy (Whitham-Toda hierarchy) inherits various 
features of the affine Toda field hierarchy. For instance, 
it has two infinite series of flows, and the flows of 
$T_s,T_{2s},\ldots$ and $\Tbar_s,\Tbar_{2s},\ldots$ are trivial. 
Note that they cannot be deduced directly from the properties 
of the affine Toda chain system. It is not the Toda chain 
system but the Toda field system that is responsible for 
the Whitham dynamics.

\section{Conclusion}

We have shown that an isomonodromy problem underlies  
Seiberg-Witten solutions of four-dimensional supersymmetric 
gauge theories. As emphasized in Introduction, integrable 
structures of the Seiberg-Witten solutions are considerably 
involved, and several different classical integrable systems 
appear to be related to the same Seiberg-Witten solution. 
Our main observation is that those apparently different 
features of integrability can be derived from a single 
isomonodromy problem. This is not just of purely mathematical 
interest.  Quantum gravity and topological conformal field 
theories in two dimensions are all reorganized into some 
isomonodromy problems 
\cite{bib:Moore1,bib:Moore2,bib:Dubrovin-ttbar,bib:Dubrovin-TCFT}. 
We believe that our isomonodromy problem, too, will be 
essentially ``stringly'',  and somehow related to the recent 
string theoretical interpretation of the Seiberg-Witten theory 
\cite{bib:Klemm-etal,bib:Kachru-etal}. 

Such a possible link with string theories is particularly 
plausible in the case of $N=2$ SU($s$) supersymmetric 
Yang-Mills theories without matter. We have considered 
this case in some detail to illustrate our idea. The 
isomonodromy problem of this case is obtained from 
the affine SU($s$) Toda field system with a constraint 
(homogeneity condition).  In principle, the other classical 
gauge groups can be treated in the same way.  We have noticed 
that this isomonodromy problem is implicitly related, via the 
third Painleve equation and its multicomponent analogues,  
to $t\tbar$ fusion of topological sigma models. 

A technical clue of our calculations is the method of 
multiscale analysis (a refined version of the ordinary 
WKB analysis) . This method enables us, just as in the 
analysis of two-dimensional quantum gravity, to derive 
the hypothetical picture of the Seiberg-Witten 
solutions as ``adiabatic deformation of finite-band 
solutions in a soliton equation''.  A mathematically 
rigorous proof will, however, require more careful and 
hard analytical considerations.  (A similar problem is 
also recently studied by mathematicians from a different
point of view \cite{bib:Painleve-multiscale}.)  

The next interesting targets are $N=4$ supersymmetric 
Yang-Mills theories. We expect that this will lead to an 
isomonodromy problem over an elliptic curve. More general Hitchin 
systems over a complex algebraic curve of arbitrary genus, too, 
will be similarly converted to an isomonodromy problem. This will  
be a nice way to construct nontrivial isomonodromy problems 
on a Riemann surface. Actually, isomonodromy problems over 
Riemann surfaces are already studied by mathematicians in 
a more general context \cite{bib:isomonodromy-anygenus}. 
Multiscale analysis of those generalized isomonodromy problems 
is mathematically a tough issue, but it deserves to be studied.


\newpage

\begin{center}
\begin{Large}
\bf Corrections to the First Version 
\end{Large}
\end{center}

\bigskip
\bigskip

\begin{description}

\item{6th line in ABSTRACT:}\newline
``(adiabatic deformation of an isomonodromy problem)''
$\longrightarrow$ 
``(adiabatic deformation of an isospectral problem)''

\item{Three items in 5th paragraph of section 1:}\newline
``1. A inite'' $\longrightarrow$ ``1. A finite''

\item{4th line in the 6th paragraph of section 1:}\newline
``adiatatic deformation'' $\longrightarrow$ ``adiabatic deformations''

\item{4th line above eq. (1):}\newline
``isomonodromic deformations.'' $\longrightarrow$ 
``isospectral deformations.''

\item{1st line below eq. (10):}\newline
``(or matrix-valued)'' $\longrightarrow$ ``(or matrix-)valued''

\item{eq. (13):}\newline
Insert ``$\Bigl( \rd^2 S(z) / \rd z^2 \Bigr)^{-1/2}$'' 
in front of ``$\exp$''

\item{eq. (21):}\newline
Insert ``$\Bigl( \rd^2 S(T,z) / \rd z^2 \Bigr)^{-1/2}$'' 
in front of ``$\exp$''

\item{2nd-3rd lines in the last paragraph of section 5:}\newline
``This lead to an isomonodromy problem over an elliptic curve. 
More general Hitchin systems over a complex algebraic curve of 
arbitrary genus, too, can be $\ldots$''
$\longrightarrow$
``We expect that this will lead to an isomonodromy problem 
over an elliptic curve. More general Hitchin systems over a 
complex algebraic curve of arbitrary genus, too, will be $\ldots$''

\end{description}

\end{document}